\documentclass[prl,a4paper,twocolumn,groupedaddress,english,showpacs]{revtex4-1} 
\usepackage{babel}
\usepackage{xcolor} 
\usepackage{graphicx} 
\usepackage{amsmath} 
\usepackage{amssymb}
\def\be{\begin{equation}} \def\ee{\end{equation}}

\def\bea{\begin{eqnarray}} \def\eea{\end{eqnarray}}

\definecolor{darkblue}{rgb}{0.1,0.2,0.6} \definecolor{darkred}{rgb}{0.8,0.1,0.2}
\usepackage[colorlinks,citecolor=darkblue,linkcolor=darkred,urlcolor=darkblue]{hyperref}
\usepackage[all]{hypcap}

\makeatletter \adddialect\l@English\l@english \makeatother

\newcommand{\bra}[1]{\langle\,#1\,|} \newcommand{\ket}[1]{|#1\rangle}

\begin{document}
\title{Many-body localization edge in the random-field Heisenberg chain} \author{David J. Luitz}
\affiliation{Laboratoire de Physique Th\'eorique, IRSAMC, Universit\'e de Toulouse, {CNRS, 31062
Toulouse, France}} 
\email{luitz@irsamc.ups-tlse.fr} \email{laflo@irsamc.ups-tlse.fr} \email{alet@irsamc.ups-tlse.fr}
\author{Nicolas Laflorencie}
\affiliation{Laboratoire de Physique Th\'eorique, IRSAMC, Universit\'e de Toulouse, {CNRS, 31062
Toulouse, France}} 
\author{Fabien Alet} \affiliation{Laboratoire de Physique Th\'eorique, IRSAMC,
    Universit\'e de Toulouse, {CNRS, 31062 Toulouse, France}} 
\date{November 3, 2014}

\begin{abstract} We present a large scale exact diagonalization study of the one dimensional spin
    $1/2$ Heisenberg model in a random magnetic field. In order to access properties at
    varying energy densities across the entire spectrum for system sizes up to $L=22$ spins, we use
    a spectral transformation which can be applied in a massively parallel fashion. Our results
    allow for an energy-resolved interpretation of the many body localization transition including
    the existence of an extensive many-body mobility edge.  The ergodic phase is well characterized by
    Gaussian orthogonal ensemble statistics, volume-law entanglement, and a full delocalization in the Hilbert space. Conversely,
    the localized regime displays Poisson statistics, area-law entanglement and 
    non ergodicity in the Hilbert space where a true localization never occurs. We perform finite
    size scaling to extract the critical edge and exponent of the localization length divergence.  
    \end{abstract} 
    \pacs{75.10.Pq, 72.15.Rn, 05.30.Rt}
\maketitle
%
%%%%%%%%%%%%%%%%%%%%%%%%%%%%%%%%
The interplay of disorder and interactions in quantum systems can lead to several intriguing
phenomena, amongst which the so-called many-body localization has attracted a huge interest in
recent years. Following precursors
works~\cite{fleishman_interactions_1980,altshuler_quasiparticle_1997,jacquod_emergence_1997,georgeot_integrability_1998},
perturbative calculations~\cite{gornyi_interacting_2005,basko_metalinsulator_2006} have established that the celebrated Anderson localization~\cite{anderson_absence_1958} can survive
interactions, and that for large enough disorder, many-body eigenstates can also ``localize'' (in a
sense to be precised later) and form a new phase of matter commonly referred to as the many-body
localized (MBL) phase.

The enormous boost of interest for this topic over the last years can probably be ascribed to the
fact that the MBL phase challenges the very foundations of quantum statistical physics, leading to
striking theoretical and experimental consequences~\cite{review_nandkishore,review_altman}. 
Several key features of the MBL phase can be highlighted as follows. It is non-ergodic, and breaks
the eigenstate thermalization hypothesis
(ETH)~\cite{deutsch_quantum_1991,srednicki_chaos_1994,rigol_thermalization_2008}: a closed system in
the MBL phase does not thermalize solely following its own dynamics. The possible presence of a
many-body mobility edge (at a finite energy density in the spectrum) indicates that conductivity
should vanish in a finite temperature range in a MBL system~\cite{gornyi_interacting_2005,basko_metalinsulator_2006}.
Coupling to an external bath will eventually destroy the properties of the MBL phase, but recent
arguments show that it can survive and be detected using spectral signatures for weak
bath-coupling~\cite{nandkishore_spectral_2014}. This leads to the suggestion that the MBL phase can
be characterized experimentally, using e.g. controlled echo experiments on reasonably well-isolated
systems with dipolar
interactions~\cite{serbyn_interferometric_2014,kwasigroch_bose_2014,yao_many_2013,vasseur_quantum_2014}.
Another appealing aspect (with experimental consequences for self-correcting memories) is that MBL
systems can sustain long-range, possibly topological, order in situations where equilibrated systems
would
not~\cite{huse_localization_2013,bahri_localization_2013,chandran_many_2014,bauer_area_2013,vosk_dynamical_2014}.
Finally, a striking phenomenological approach~\cite{huse_phenomenology_2014} pinpoints that the MBL
phase shares properties with integrable systems, with extensive local integrals of
motion~\cite{serbyn_local_2013,ros_integrals_2014,chandran_constructing_2014}, and that MBL
eigenstates sustain low (area law) entanglement. This is in contrast with eigenstates at finite
energy density in a generic equilibrated system, which have a large amount (volume law) of
entanglement and which are believed to be well described within a random matrix theory approach.
%%%%%%%%%%%%%%%%%%%%%%%%%%%%%%%%
    \begin{figure}[t] \centering \includegraphics[width=\columnwidth,clip]{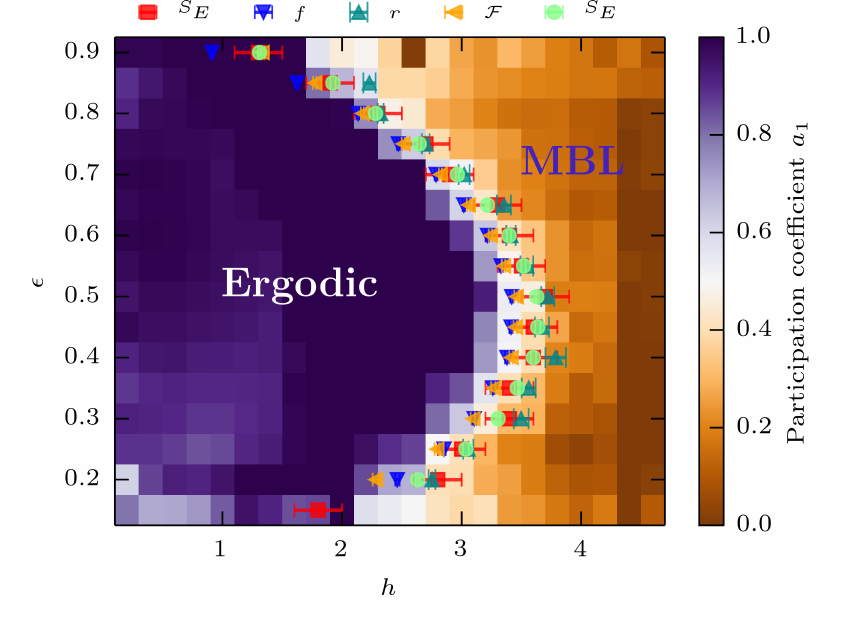}
        \vspace{-1.cm} \caption{Disorder ($h$) --- Energy density ($\epsilon$) phase diagram of the
            disordered Heisenberg chain Eq.~\eqref{eq:H}. The ergodic phase (dark region with a
            participation entropy volume law coefficient $a_1\simeq 1$) is separated from the
            localized regime (bright region with $a_1 \ll 1$). Various symbols (see legend) show the
        energy-resolved MBL transition points extracted from finite size scaling
        performed over system sizes $L\in\{14,15,16,17,18,19,20,22\}$. Red squares correspond to a visual
estimate of the boundary between volume and area law scaling of entanglement entropy $S_E$.}
\label{fig:a1phasediag} \end{figure}
%%%%%%%%%%%%%%%%%%%%%%%%%%%%%%%%

Going beyond perturbative approaches, direct numerical simulations of disordered quantum interacting
systems provide a powerful framework to test MBL features in a variety of
systems~\cite{oganesyan_localization_2007,pal_many-body_2010,znidaric_many_2008,cavoni_quantum_2011,cuevas_level_2012,bauer_area_2013,serbyn_interferometric_2014,kjall_many-body_2014,luca_ergodicity_2013,Iyer_many_2014,pekker_hilbert_2014,johri_numerical_2014,bardarson_unbounded_2012,andraschko_purification_2014,laumann_many_2014,vasseur_quantum_2014,hickey_signatures_2014,nanduri_entanglement_2014,barlev_dynamic_2014}.
The MBL transition dealing with eigenstates at high(er) energy, ground-state methods are not well
adapted. Most numerical studies use full exact diagonalization (ED) to obtain all eigenstates and
energies and are limited to rather small Hilbert space sizes $\dim{\cal H} \sim 10^4$~\footnote{We
    note that dynamics after a quench can be investigated on fairly large systems using methods that
    benefit from the low rise of entanglement in a MBL
    system~\cite{znidaric_many_2008,bardarson_unbounded_2012,andraschko_purification_2014}}. 
    
    In this Letter, we present an extensive numerical study of the periodic $S=\frac{1}{2}$ Heisenberg
    chain in a random magnetic field, governed by the Hamiltonian
\be H=\sum_{i\in [1,L]} {\bf S}_i \cdot {\bf  S}_{i+1} -h_iS_i^z, \label{eq:H} \ee
with $h_i$ drawn from a uniform distribution $[-h,h]$ (total magnetization $S^z$ is
conserved). Model~\eqref{eq:H} has been
used~\cite{pal_many-body_2010,bauer_area_2013,luca_ergodicity_2013,nanduri_entanglement_2014} as a
prototype for the MBL transition in the ``infinite-temperature'' limit, where the full many-body
spectrum (or a large fraction thereof) is considered for systems of maximum size $L\approx 16$. In
this work, we instead use a shift-inverse ED approach and are able to reach eigenstates at arbitrary
energy density for systems up to $L=22$ with very large Hilbert spaces ($ \dim{{\cal
H}_{L=22}}=705\,432$ in the $S^z=0$ sector). Our simulations unambiguously reveal the existence of
an extensive
many-body localization edge: the resulting phase diagram (disorder strength $h$ {\it{vs.}} energy
density $\epsilon$, Fig.~\ref{fig:a1phasediag}) is built on a careful finite size scaling analysis
of numerous energy-resolved estimates. In particular the transition is captured using, {\it{e.g.}}
spectral statistical correlations between nearby eigenstates, volume {\it{vs.}} area law of
entanglement entropies and bipartite fluctuations, spin relaxation, localization properties in the
Hilbert space, which all roughly agree within error
bars. We also perform a scaling analysis close to the MBL transition.
 
{\it{Characterization of ergodic and localized regimes---}} Before presenting our numerics, we
summarize the main differences between ergodic and localized phases, and the observables used to quantify
them. 

\begin{figure}[t] \centering \includegraphics{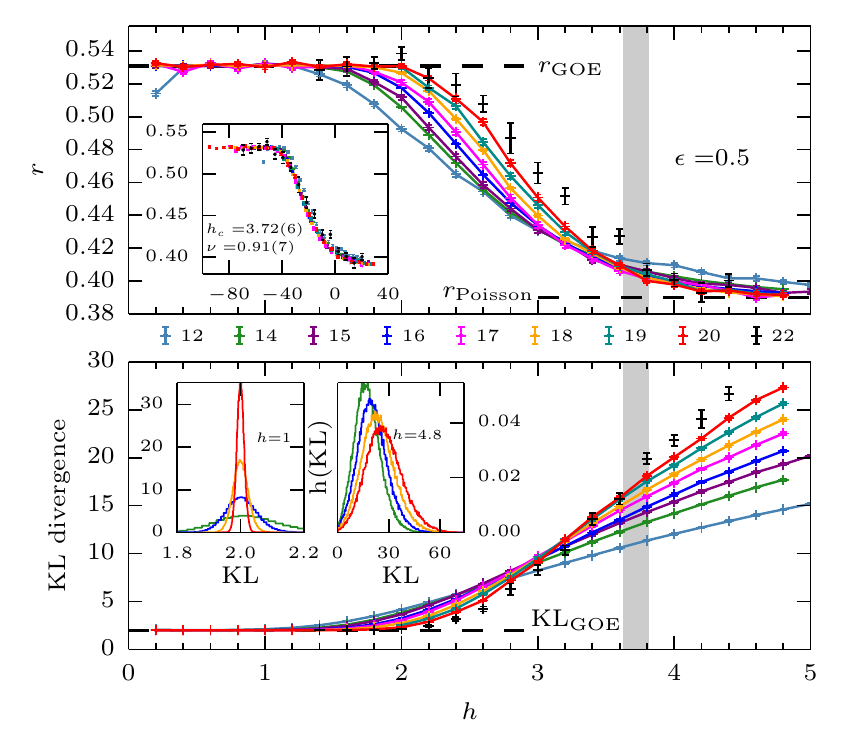} \vspace{-1.cm} \caption{Adjacent gap
    ratio (top) and Kullback Leibler divergence (bottom) as a function of disorder strength in the
spectrum center $\epsilon=0.5$. Insets: (top) data collapse used to extract the critical disorder
strength $h_c$ and exponent $\nu$. The $h$ axis is transformed by $(h-h_c) L^{1/\nu}$,
(bottom) distribution of KLd in both phases.} \label{fig:kl_r}
\end{figure}

{\it{(i) Level statistics and eigenvectors similarity. }} A popular way to differentiate extended and localized regimes relies on studying spectral statistics using tools from random
matrix theory~\cite{rmt}. In the ergodic regime, the statistical distribution of level spacings
follows Wigner's surmise of the Gaussian orthogonal ensemble (GOE), while a Poisson distribution is
expected for localized states. It is convenient~\cite{oganesyan_localization_2007} to consider the
ratio of consecutive level spacings
$r^{(n)}={\rm{min}}(\delta^{(n)},\delta^{(n+1)})/{\rm{max}}(\delta^{(n)},\delta^{(n+1)})$ with
$\delta^{(n)}=E_n-E_{n-1}$ at a given eigen-energy $E_n$ to discriminate between the two phases, as
its disorder average changes from ${{r}}_{\rm GOE}=0.5307(1)$~\cite{atas_distribution_2013}
to ${{r}}_{\rm Poisson}=2\ln 2-1\simeq 0.3863$. This has been used in several
works~\cite{oganesyan_localization_2007,pal_many-body_2010,cuevas_level_2012,johri_numerical_2014,laumann_many_2014,bauer_area_2013},
averaging over a large part of the  spectrum. Here, we compute ${{r}}$ in an
energy-resolved way in order to locate the MBL edge (Fig.~\ref{fig:kl_r}). 

Quite interestingly, the GOE--Poisson transition can also be captured by correlations between nearby
eigenstates. We expect eigenfunctions to be ``similar'' (``different'') in the ergodic (localized)
regime. We quantify the degree of correlation by the Kullback-Leibler divergence
(KLd)~\cite{kullback1951}, defined by ${\rm KL}=\sum_{i=1}^{\dim{\cal H}}p_i\ln(p_i/q_i)$, where
$p_i=|\langle i | n \rangle|^2$ and $q_i=|\langle i | n' \rangle|^2$ are the moduli squared of the
wave functions coefficients of 2 nearby eigenstates $|n\rangle, | n' \rangle$ expressed in the
computational basis $\{ |i\rangle \}$ (here  $\{S^z\}$). The KLd displays different behavior in
the two phases (Fig.~\ref{fig:kl_r}): we find ${\rm{KL}}_{\rm GOE}=2$~\footnote{We found this 
numerically for eigenvectors of random matrices in the GOE ensemble, inspiring an analytical proof (O. Giraud, private communication).}, 
and ${\rm{KL}}_{\rm Poisson}\sim \ln (\dim{\cal
H})$.

\begin{figure}[b] \centering \includegraphics{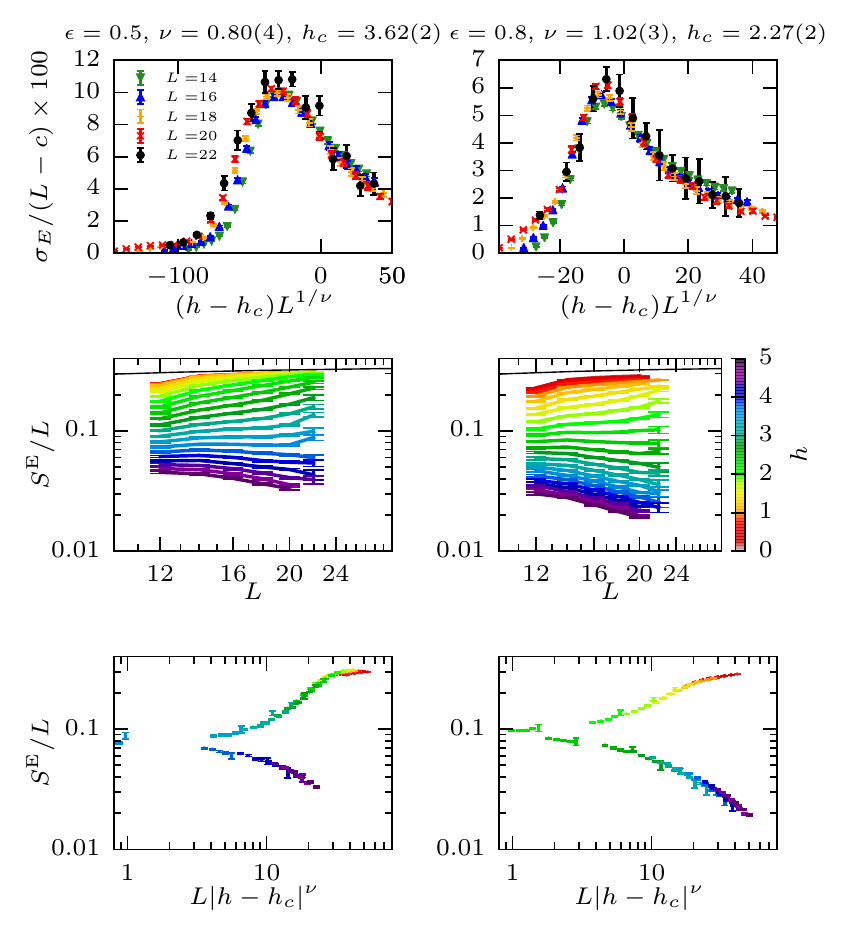} \vspace{-1.cm}
    \caption{Entanglement entropy per site $S^\text{E}/L$ and its variance $\sigma_{E}$, as a
    function of system size $L$ for different disorder strengths in the middle of the spectrum
    (left) and in the upper part (right). The volume law scaling leading to a constant
    $S^\text{E}/L$ for weak disorder contrasts with the area law (signaled by a decreasing
    $S^\text{E}/L$) at larger disorder is very clear. Black line: $S^\text{E}/L$ for a random state 
    \cite{page_average_1993}.  Close to the transition, the prefactor of
the volume law is expected to converge only for larger system sizes.} \label{fig:EEvsL} \end{figure}

\begin{figure}[t] \centering \includegraphics{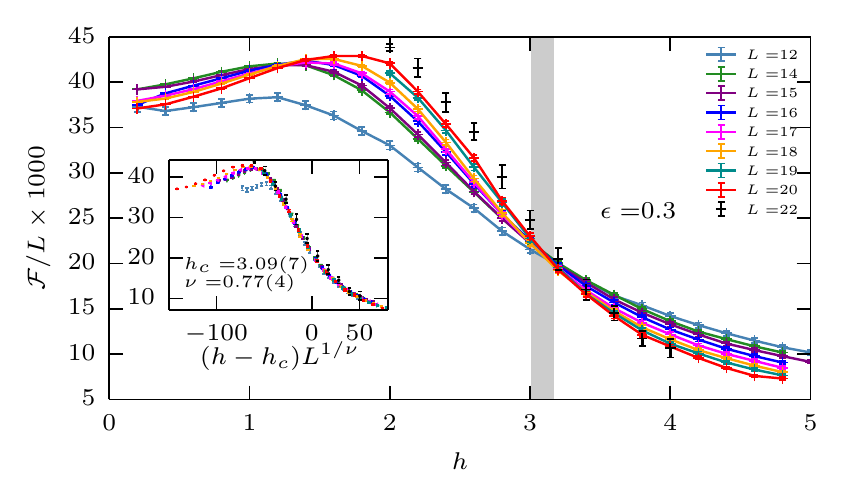} \vspace{-1.cm} \caption{ Bipartite
        fluctuations of half-chain magnetization as a function of disorder strength at
        $\epsilon=0.3$. Inset: data collapse using the best estimates for the critical
        disorder strength $h_c=3.09(7)$ and $\nu=0.77(4)$.  } \label{fig:bf} \end{figure}

{\it{(ii) Entanglement entropy (EE). }} Beyond level statistics, EE provides a quantitative tool to
characterize how information is spread from one part of the system to an
other~\cite{review_nandkishore}.  In the ergodic regime satisfying the ETH, the reduced density
matrix $\rho_A$  of a typical eigenstate is expected to be thermal, yielding a volume-law scaling
(with the subsystem $A$ size) for the entanglement entropy $S^E=- {\rm Tr} \rho_A \ln \rho_A$.
Conversely, localized eigenstates display a much smaller entanglement, expected to cross-over
towards an area-law scaling~\cite{review_nandkishore,bauer_area_2013} when the subsystem size
exceeds the localization length. These different scalings of $S^E$ allow to distinguish both regimes
(Fig.~\ref{fig:EEvsL}). In the same spirit, we expect bipartite fluctuations of the subsystem
magnetization $S^z_A$~\cite{song_bipartite_2012} $\mathcal{F}=\langle \left( S^z_A \right)^2 \rangle
- \langle S^z_A \rangle^2$ to exhibit similar scaling (Fig.~\ref{fig:bf}).

\begin{figure}[b] \centering \vspace{-0.5cm} \includegraphics{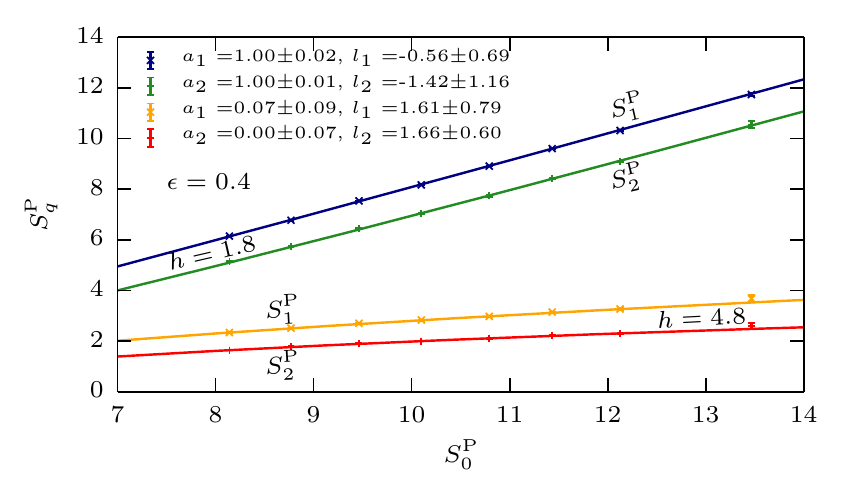} \vspace{-1.cm}
        \caption{Participation entropy as a function of $S_0^P=\ln(\text{dim}\mathcal{H})$ for $q=1, 2$ and $\epsilon=0.4$. In the ergodic phase ($h=1.8$),
            $S_q^P$ grows linearly with $S_0^P$ while the linear scaling term vanishes within our
            error bars in the localized regime ($h=4.8$). Our fits (solid lines, see text) constrain $a_q\in[0,1]$ and yield a logarithmic scaling prefactor $l_q\approx 2(1)$ at $h=4.8$, consistent with a (slow) growth of $S_q^P$ with system
            size  in the localized phase.
        } 
            \label{fig:pe}
    \end{figure}

{\it{(iii) Hilbert space localization.}} Another characterization of MBL relies on inverse participation ratios and associated participation
entropies (PE), traditionally used in the context of single particle
localization~\cite{bell_dynamics_1972,wegner_inverse_1980,rodriguez_multifractal_2011} and recently
for many-body physics~\cite{luitz_universal_2014,luitz_participation_2014}. Here the localization is studied in the Hilbert
space (of dimension $\dim{\cal H}$) of spin configurations via the disorder average PEs
${S^\text{P}_q}$, defined for any eigenstate $\ket{n}$ represented in the $\{S^z\}$  basis by
$S^\text{P}_q(\ket{n}) = \frac{1}{1-q}\ln\sum_i p_i^{q}$ [$S^\text{P}_1(\ket{n})=-\sum_i p_i \ln p_i$]. We generically find eigenstates
    to be delocalized  {\it{in both}} regimes with qualitatively different features. In the ergodic
    regime, we obtain a leading scaling $S^{\rm P}_q=a_q\ln ( { \dim{\cal H} })$ with $a_q\approx 1$
    $\forall q$ (see color coding of $a_1$ in Fig.~\ref{fig:a1phasediag}). In the localized phase,
    PE also grows with system size (Fig.~\ref{fig:pe}), but much slower with $a_q\ll 1$, or $a_q =0$
    within error bars and a slow log divergence $S^{\rm P}_q = l_q \ln ( \ln{ \dim{\cal H} } )
    $, indicating a non-trivial multi-fractal behavior.
    
\textit{Numerical method ---} 
The complete diagonalization of the non-translation invariant
Hamiltonian Eq.~\eqref{eq:H} is out of reach for system sizes $L \gtrsim18$ spins. 
Therefore, we use an approach successful for the Anderson localization problem (see e.g.
Ref.~\onlinecite{rodriguez_multifractal_2011}) and restrict ourselves to certain energy slices in
the spectrum by using a shift-invert spectral transformation $(\mathbf{H}-E\mathbf{1})^{-1}$. In the transformed problem, it is easy to apply Krylov space
methods~\cite{hernandez_slepc_2005} to compute the eigenpairs closest to the shift energy $E$.

For each disorder realization, we first calculate the extremal
eigenenergies $E_0$ and $E_\text{max}$ used to define the normalized energy target $\epsilon =
(E-E_\text{max})/(E_0-E_\text{max})$ (we considered the $S^z=0$ sector of even-sized
$L=12,14,16,18,20,22$ and $S^z=1$ sector of $L=15,17,19$). The shift-invert method, based on a
massively parallel $LU$ decomposition~\cite{MUMPS1,MUMPS2}, is then used to calculate at least 50
eigenpairs with energy densities closest to the targets $\epsilon=\{0.05,0.1,\dots 0.95\}$. Note that this
is a much more demanding computational task than for the Anderson problem, as the number of
off-diagonal elements of $H$ scales with $L$. We use at least 1000 disorder realizations for each
$L$ (except for $L=22$ where we accumulated between 50 and 250 samples). For each $\epsilon$, observables
are calculated from the corresponding eigenvectors and averaged over target packets and disorder
realizations for each value of the disorder strength $h$. As eigenvectors of the same disorder
realization are correlated, we found it crucial~\cite{rodriguez_multifractal_2011} to bin quantities
over all eigenstates of the same realization, and then compute the standard error over these bin averages, in order not to underestimate error bars. Investigating numerous
quantities allows to check the consistency of our analysis and conclusions.

%%%%%%%%%%%%%%%%%%%%%%%%%%%%%%%%%%
%%%%%%%%%%%%%%%%%%%%%%%%%%%%%%%%%%
%%%%%%%%%%%%%%%%%%%%%%%%%%%%%%%%%%

\noindent\textit{Results and finite size scaling analysis---} We discuss the transition between GOE
and Poisson statistics, first using the consecutive gap ratio ${\overline{r}}$, shown in
Fig.~\ref{fig:kl_r} (top) for $\epsilon=0.5$.  When varying the disorder strength
$h$, we clearly see a crossing around $h_c\sim 3.7$ between the two limiting values. This crossing
can be analyzed using a scaling form $g[L^{1/\nu}(h-h_c)]$ which allows a collapse of the data onto a single universal curve (see
    inset), yielding $h_c=3.72(6)$ and $\nu=0.91(7)$ (see details of fitting procedure and error bars estimates in Sup. Mat.). 

The above defined KLd, computed for two eigenstates randomly chosen at the same energy target
$\epsilon$ and averaged over disordered samples, also displays a crossing between the two limit
scalings ${\rm{KL}}_{\rm GOE}=2$ and ${\rm{KL}}_{\rm Poisson}\sim \ln (\dim{\cal H})$
(Fig.~\ref{fig:kl_r} bottom). A data collapse is very difficult to achieve for KL due to a large
drift of the crossing points. Nevertheless, the distributions of KL plotted in insets, display
markedly different features. The perfect gaussian distribution in the ergodic phase (at $h=1$)
around the GOE mean value of 2 with a variance decreasing with $L$ provides strong evidence that the
statistical behavior of the \emph{eigenstates} is well described by GOE, extending its applicability
beyond simple level statistics. In the MBL regime ($h=4.8$), the behavior is completely different as variance and
mean both increase with $L$.

%%%%%%%%%%%%%%%%%%%%%%%%%%%%%%%%%%
%%%%%%%%%%%%%%%%%%%%%%%%%%%%%%%%%%

We now turn to the entanglement entropy for a real space bipartition at $L/2$ ($L$ even).
Shown for two targets $\epsilon=0.5$ and $0.8$, the transition is signaled (Fig.~\ref{fig:EEvsL}) by
a change in the EE scalings from volume law $S^{\rm E}/L\to {\rm constant}$ for $h< h_c$ to area-law
with $S^{\rm E}/L\to 0$ for $h>h_c$. Assuming a volume law scaling at the critical
point~\cite{grover_certain_2014}, we perform a collapse of $S^{\rm E}/L$ to the form
$g[L^{1/\nu}(h-h_c)]$ (Fig.~\ref{fig:EEvsL} bottom panel) giving estimates for the critical disorder
$h_c$ and exponent $\nu$ consistent with other results (see Sup. Mat.). Furthermore, as recently
argued~\cite{kjall_many-body_2014}, the standard deviation of the entanglement entropy displays a
maximum at the MBL transition. A scaling collapse of the form $\sigma_E=(L-c) g[L^{1/\nu}(h-h_c)]$
(with $c$ an unknown parameter and the previous estimates of $\nu$ and $h_c$ from collapse of $S^{\rm E}/L$) works particularly well (top panel of Fig.~\ref{fig:EEvsL}).

Perhaps more accessible to experiments, bipartite fluctuations ${\cal F}$ of subsystem magnetization (taken here to be a half-chain $L/2$) have a similar behavior. Being
simply the Curie constant of the subsystem, we also expect thermal extensivity (subextensive
response) in the ergodic (localized) regime.  This is clearly checked in Fig.~\ref{fig:bf} for
$\epsilon=0.3$ where ${\cal F}/L$ has a crossing point at the disorder-induced MBL transition. A
data collapse (inset of Fig.~\ref{fig:bf}) is also possible for ${\cal F}/L=g[L^{1/\nu}(h-h_c)])$,
giving $h_c=3.09(7)$ and $\nu=0.77(4)$, consistent with estimates from other quantities
(Fig.~\ref{fig:a1phasediag}). Finally, we also performed an analysis of the dynamic fraction
$f$ of an initial spin polarization~\cite{pal_many-body_2010}, and obtained similar
consistent scaling (see Supp. Mat. and Fig.~\ref{fig:a1phasediag}).

%%%%%%%%%%%%%%%%%%%%%%%%%%%%%%%%%%
%%%%%%%%%%%%%%%%%%%%%%%%%%%%%%%%%%

%%%%%%%%%%%%%%%%%%%%%%%%%%%%%%%%%%
%%%%%%%%%%%%%%%%%%%%%%%%%%%%%%%%%%

The disordered many-body system can be mapped onto a single particle problem on the complex
graph spanned by the Hilbert space whose $\dim{\cal H}$ vertices are the basis states, which are
connected by spin-flip terms in Eq.~\eqref{eq:H}. The average coordinance of each node is $z\sim L$
and the random potential has a gaussian distribution of variance $\sigma_h\sim h\sqrt L$, meaning
that the effective connectivity grows faster than the disorder strength. Using recent results on
Anderson localization on Bethe lattices at large connectivity~\cite{biroli_anderson_2010}, we do not
expect genuine Hilbert space localization at any finite disorder. This argument is corroborated by our
numerical results for the PE $S_q^{\rm P}$ (Fig.~\ref{fig:pe}) which are always found to increase
with $S_0^{\rm P} \equiv \ln(\dim{\cal H})$, albeit much more slowly in the localized regime. Analysis
of various fits of the form  $S_q^{\rm P}=a_q S_0^{\rm P}+l_q \ln (S_0^{\rm P})+o(S_0^{\rm P})$
indicate that $a_q \simeq 1$  $\forall q$ in the ergodic regime (with possibly small negative $l_q$
corrections) as seen in the color scale of Fig.~\ref{fig:a1phasediag}, in contrast to Anderson
localization on the Bethe lattice~\cite{de_luca_anderson_2014}. In the localized regime, we
obtain essentially similar fit qualities with $a_q \ll 1$ (see typical numbers in
Fig.~\ref{fig:pe}), or $a_q=0$ and $l_q>0$ (the slow growth of  $S_q^{\rm P}$ and our limited system
sizes do not allow to separate these two possibilities).

{\it{Discussions and conclusions---}} Using various estimates for the MBL transition, our
large-scale energy-resolved ED results indicate the existence of an extensive many-body mobility edge in the excitation spectrum (Fig.~\ref{fig:a1phasediag}) of the random field Heisenberg chain. Furthermore, we
show that the ergodic regime has full features of a metallic phase (with $a_q=1$ and GOE statistics
for both energy levels \emph{and} the wavefunction coefficients), and that the localized
many-body states do not exhibit a true Hilbert-space localization for configuration spaces
up to $\dim{\cal H} \sim 7\cdot 10^5$\footnote{We cannot exclude a different multifractal behavior
$0<a_q<1$ very close to the transition}. Our detailed finite-size scaling analysis (Sup. Mat.)
provides a consistent estimate of
a characteristic length diverging as $|h-h_c|^{-\nu}$ with $\nu=0.8(3)$ through the full phase
diagram. This estimate of the exponent $\nu$ appears to violate the Harris-Chayes~\cite{harris_effect_1974,chayes_finite_1986} criterion $\nu
\geq 2/d$ (see also Ref.~\cite{kjall_many-body_2014}) within the system sizes used. This is quite intriguing given that for the same size range, the location of the critical point is consistent for all various estimates used (see Fig.~\ref{fig:a1phasediag}). This opens new questions on the finite-size scaling and/or corrections to scaling at the MBL transition which may not follow~\cite{oganesyan_localization_2007,pal_many-body_2010} standard forms.

Besides these results for the particular model Eq.~\ref{eq:H}, we believe that the numerical
techniques (massively parallel energy-resolved diagonalisation) and new indicators of the
ergodic-localized transition (eigenstates correlations or bipartite fluctuations) introduced here
will be useful in a large number of contexts related to MBL or ETH. In particular, the obtention of
exact eigenvectors on fairly large systems will be crucial to quantify the effectiveness of encoding
localized states as matrix product states, as recently
advocated~\cite{friesdorf_many_2014,chandran_spectral_2014,pekker_encoding_2014}.

{\it{Acknowledgments ---}}  We thank G. Lemari\'e, F. Pollmann, B. Georgeot, O. Giraud for fruitful
discussions, and CALMIP for generously providing access to the EOS
supercomputer. We used the libraries PETSc, SLEPc~\cite{hernandez_slepc_2005} and the
MUMPS~\cite{MUMPS1,MUMPS2} parallel solver for our calculations. This work was performed using HPC
resources from GENCI (grant x2014050225) and CALMIP (grant 2014-P0677), and is supported by the
French ANR program ANR-11-IS04-005-01.

\vspace{-0.57cm}
\bibliography{mbl}

\appendix

\section{Supplementary material}

\subsection{Details on fitting procedures and estimates of critical exponents and fields}

\begin{figure}[h]
    \centering
    \includegraphics{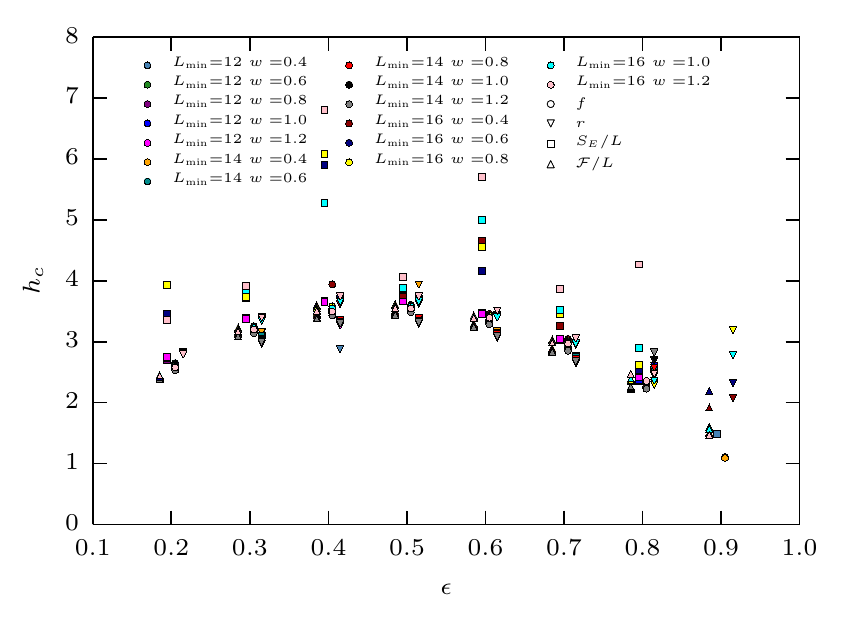}
    \caption{Systematic analysis of the influence of fit windows in $h$ and $L$ on the critical
    disorder strength $h_c$. Only results for the targets $0.1$, 0.2, 0.3, 0.4, 0.5, 0.6, 0.7, 0.8
and 0.9 are shown and the symbols were slightly shifted in $\epsilon$ for better readability.}
    \label{fig:sys_hc}
\end{figure}

\begin{figure}[h]
    \centering
    \includegraphics{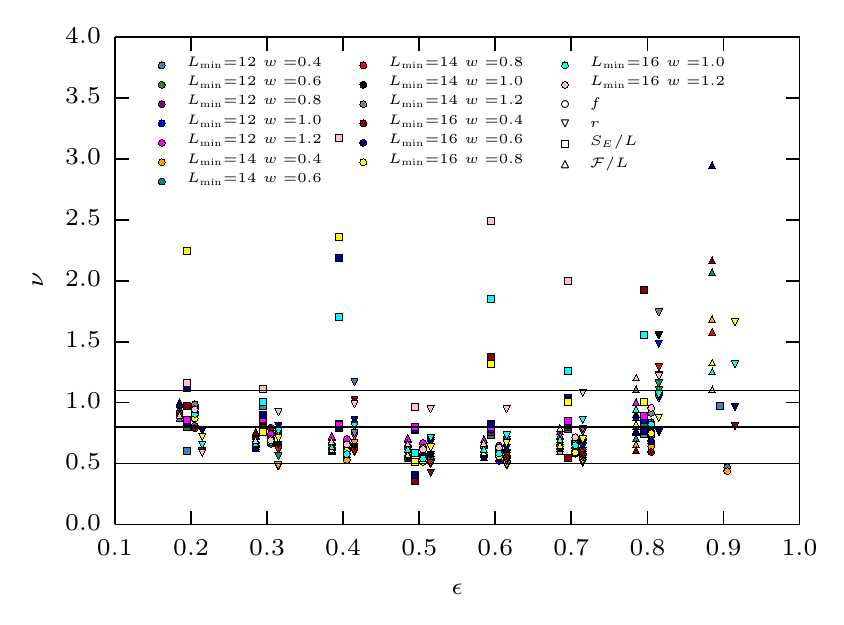}
    \caption{Systematic analysis of the influence of fit windows in $h$ and $L$ on the critical
    exponent $\nu$, the horizontal lines correspond to the mean value and the error bounds of our
estimate for the critical exponent $\nu=0.8(3)$. Only results for the targets $0.1$, 0.2, 0.3, 0.4, 0.5, 0.6, 0.7, 0.8
and 0.9 are shown and the symbols were slightly shifted in $\epsilon$ for better readability.}
    \label{fig:sys_nu}
\end{figure}

In order to estimate the value of the critical disorder strength $h_c$ and the critical exponent
$\nu$, we have performed a systematic scaling analysis using the scaling ansatz $g[ (h-h_c)
L^{1/\nu}]$ for the disorder averaged gap ratios ${r}$, the dynamical spin fraction ${f}$,
the entanglement entropy per site $S_\text{E}/L$ and the bipartite fluctuations per site
$\mathcal{F}/L$. We model the universal function $g$ in a window of size $2w$ centered at $h_c$ by a
polynomial of degree three and have performed fits varying the size of the fit window and excluding
system sizes smaller than $L_\text{min}$ for $L_\text{min}\in\{12,14,16\}$ in order to estimate the
stability of our results. We have also tried to include drift terms in the universal function but
concluded that they are not needed to obtain a very good fit quality. The results of our stability
analysis is displayed in Figures \ref{fig:sys_hc} and \ref{fig:sys_nu}, where we show the results
of scaling fits for all quantities and fit windows. The scattering of the results can be understood
as a measure of the true error bar.

Generally, our results are all consistent and nearly all of the outliers stem from fits including
only the largest system sizes $L\geq 16$, where the analysis starts to become difficult due to the
reduced range in $L$. In particular, the analysis for the entanglement entropy per site is
problematic in this case, as we only use even system sizes.

Additionally, at the low and high end of the spectrum, the density of states is very low, thus
rendering the analysis of the gap ratios particularly problematic \cite{oganesyan_localization_2007}. 

Based on this stability analysis, we find that the fit window with $L_\text{min}=14$ (for
the gap ratios, we use $L_\text{min}=15$) and $w=0.8$ seems to provide the most stable results and is therefore used for
all results presented in the rest of this Letter. With the fixed fit window, we have performed a
bootstrap analysis in order to estimate the statistical error of the fit parameters, in particular
$h_c$ and $\nu$, indicated in the plots. Clearly, one has to keep in mind that on top of this error,
there will be a systematic error that is of the order of the spread of the results in the stability
analysis shown in this paragraph.

\subsection{Dynamical spin fraction}

\begin{figure}[b] \includegraphics{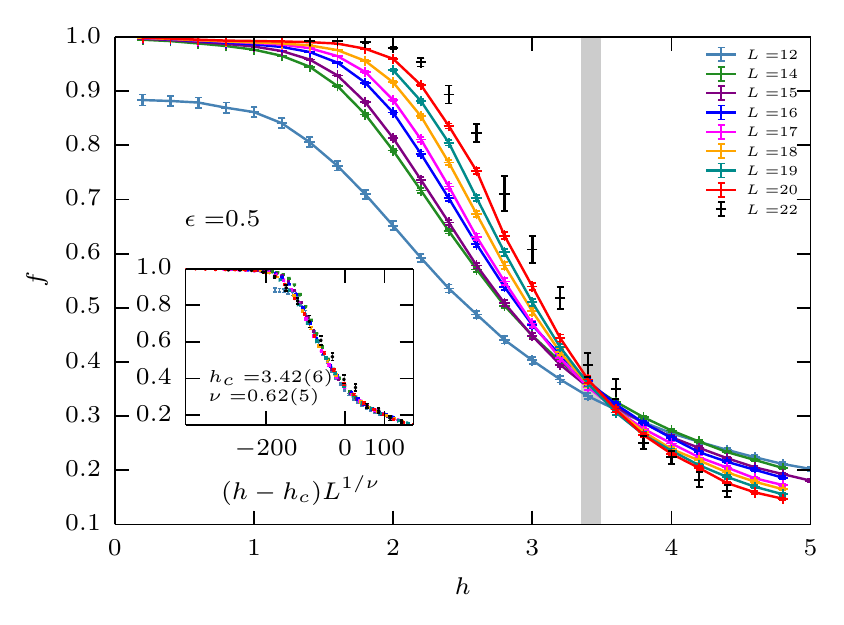} \vspace{-1.cm} \caption{${f}$ as a function
of disorder strength for different system sizes $L$ in the center of the spectrum $\epsilon=0.5$. }
\label{fig:f} \end{figure}

For completeness, we show here additional data for the dynamical spin fraction $f$, which has been
introduced in Ref.~\onlinecite{pal_many-body_2010}. This quantity gives a measure of how
much memory of an initial spin density is lost after a long time evolution. It is
$1$ (corresponding to no memory) in the ergodic phase and decays to zero in the localized phase.

It can be defined by introducing an initial spin density defined by the longest wavelength operator
\be M=\sum_{j \in [1,L]} S_j^z \exp( i 2 \pi j/L).  \ee

After evaluating the long time remainder of this spin density, one finds for the dynamic fraction
for an eigenstate $\ket{n}$ \be f^{n}= 1-\frac{ \bra{n} M^\dagger \ket{n} \bra{n} M \ket{n}
}{ \bra{n} M^\dagger M \ket{n} }. \ee

Fig.~\ref{fig:f} represents the disorder-average ${f}$ as a function of disorder strength for
different system sizes $L$ in the spectrum center $\epsilon=0.5$, where a crossing point can be
observed. Assuming a finite-size scaling of the form $g[ (h-h_c)L^{1/\nu}]$ allows to collapse all data (see inset), producing best-fit values of $\nu$ and $h_c$ (see inset) compatible with other estimates (see details of fitting procedure in first part of this Sup. Mat.).

\end{document}